\documentclass{elsart}

\usepackage{natbib}
\usepackage{lscape} 
\usepackage{epsfig,makeidx}

\usepackage[version=3]{mhchem}
\makeatletter

\newcounter{reaction}
\renewcommand{\thereaction}{R\,\arabic{reaction}}
 
\newcommand\reaction@[2]{\bgroup%
 \let\theequationold\theequation%
 \def\theequation{\thereaction}%
 \begin{equation}%
   \addtocounter{equation}{-1}%
   \refstepcounter{reaction}%
   \label{reac:#2}%
   \ce{#1}%
 \end{equation}%
 \def\theequation{\theequationold}%
 \egroup%
}

\newcommand\reaction@nonumber[1]%
{\begin{equation*}\ce{#1}\end{equation*}}

\newcommand\reaction{\@ifstar{\reaction@nonumber}{\reaction@}}

\makeatother
 
\usepackage{amssymb}

\begin{document}

\begin{frontmatter}



\title{MICS Asia Phase II -- \\ Sensitivity to the aerosol module}


\author[cerea]{K.N. Sartelet\corauthref{cor}}
\corauth[cor]{Corresponding author.}
\ead{sartelet@cerea.enpc.fr}
\author[criepi]{H. Hayami}
\author[cerea]{B. Sportisse}

\address[cerea]{CEREA, 6-8 avenue Blaise Pascal, Cit\'e Descartes Champs-sur-Marne 77455 Marne la Vall\'ee Cedex 2 }
\address[criepi]{CRIEPI, Environmental Science Research Laboratory, 1646 Abiko, Abiko-shi, Chiba 270-1194, Japan.}

\begin{abstract}
In the framework of the model inter-comparison study - Asia Phase II (MICS2), where eight models are compared over East Asia, this paper studies the influence of different parameterizations used in the aerosol module on the aerosol concentrations of sulfate and nitrate in PM$_{10}$.

An intracomparison of aerosol concentrations is done for March 2001 using different configurations of the aerosol module of one of the model used for the intercomparison. Single modifications of a reference setup for model configurations are performed and compared to a reference case. These modifications concern the size distribution, i.e. the number of sections, and physical processes, i.e. coagulation, condensation/evaporation, cloud chemistry, heterogeneous reactions and sea-salt emissions.

Comparing monthly averaged concentrations at different stations, the importance of each parameterization is first assessed. It is found that sulfate concentrations are little sensitive to sea-salt emissions and to whether condensation is computed dynamically or by assuming thermodynamic equilibrium. Nitrate concentrations are little sensitive to cloud chemistry. However, a very high sensitivity to heterogeneous reactions is observed.

Thereafter, the variability of the aerosol concentrations to the use of different chemistry transport models (CTMs) and the variability to the use of different parameterizations in the aerosol module are compared. For sulfate, the variability to the use of different parameterizations in the aerosol module is lower than the variability to the use of different CTMs. However, for nitrate, for monthly averaged concentrations averaged over four stations, these two variabilities have the same order of magnitude.

\end{abstract}

\begin{keyword}

aerosol \sep variability \sep chemistry transport model \sep size distribution \sep coagulation \sep condensation \sep cloud chemistry \sep heterogeneous reactions \sep sea-salt emissions

\end{keyword}

\end{frontmatter}


\section{Introduction}

The model inter-comparison study - Asia Phase II (MICS2), which follows the MICS Phase I \citep{mics1}, aims at comparing transport and deposition of sulfur, nitrogen compounds, ozone and aerosols in East Asia. Eight models ($M_1$, ..., $M_8$) are compared for the following four periods: March 2001, July 2001, December 2001 and March 2002. 

The eight models employed for the intercomparison use the same emission data, meteorological data, boundary conditions. Note that meteorological data that are different from the standard MICS meteorological file have sometimes been used, and that the size of the domain of study may differ from domain to domain. Apart from input data and the choice of the computational domain, the variability among the models is related to numerical schemes and parameterizations used for transport, diffusion, deposition, scavenging and chemical mechanisms.
Moreover, the eight models use different parameterizations in the aerosol module.

This paper aims investigating to which extent the use of different parameterizations in the aerosol module impact the aerosol concentrations of sulfate and nitrate. An intracomparison of aerosol concentrations is done in March 2001 using different configurations of the aerosol module of one of the model used for the intercomparison.
The intracomparison is done using the model $M_8$, i.e. Polair3D (\citet{bs-ijep}, \citet{bs-kat1} , \citet{tombette_gloream2005}, \citet{sartelet04tokyo}), that has been designed for such multi-configuration works (see for instance \citet{mallet05uncertainty}). 
Informations about the domain of study and input data may be found in \citet{mics2}.

\citet{mics-aer} have focused on the intercomparison for aerosols, more especially on inorganic components: sulfate, ammonium and nitrate in PM$_{10}$. They show that models' predictions tend to be closer to observations for sulfate than for nitrate. Sulfate tends to be slightly underestimated in March 2001. \citet{mics-aer} found that the amount of sulfate is lower for the model $M_8$ than for $M_3$, $M_5$ and $M_7$, although these four models have comparable amount of total sulfur. This can probably be explained by the omission of aqueous chemistry in $M_8$.
Although large discrepancies exist between models, especially for nitrate concentrations, total nitrate is consistently underestimated by most models. The nitrate concentrations predicted by $M_8$ are comparable to other models, but the total nitrate (nitrate + $HNO_3$) is slightly lower, which can be due to the omission of heterogeneous reactions. In $M_8$, most of the total nitrate is in the particulate phase.

This article is organised as follows. First, the different aerosol modules used in the different models that participated to the MICS study are presented with an accent on the aerosol module used in Polair3D. Then, the sensitivity to the aerosol module is studied and compared to the sensitivity to the chemistry transport model for monthly averaged concentrations at EANET (Acid Deposition Monitoring Network in East Asia) stations and at Fukue, a remote site between Japan and China. 

\section{The aerosol modules}

\subsection{The different physical processes}

In the aerosol module of eulerian models, the size distribution is often modeled using a sectional approach (e.g. \citet{ed_madm}, \citet{ed_jacob_develop3}) or a modal approach (e.g. \citet{ed_trimod2}, \citet{sartelet-mam}). In the sectional approach, the number of sections varies from study to study, usually from 2 to 15. In the modal approach, typically from 2 to 4 modes are used. 

The aerosol composition and distribution are influenced by different physical processes such as condensation/ evaporation, which is thought as one of the major processes influencing aerosol composition, heterogeneous reactions on the surface of particles, coagulation, nucleation, cloud chemistry. 
Three approaches may be used to model condensation/ evaporation processes (\citet{ed_Capaldo_hybrid}, \citet{bs-reducaer})
\begin{itemize}
\item a dynamic approach, i.e. the mass transfer between gas and aerosol phases is explicitly taken into account,
\item a full-equilibrium approach, i.e. the dynamic modeling is replaced with an assumption of thermodynamic equilibrium between the gas and aerosol phases, 
\item an hybrid approach, i.e. full equilibrium is assumed for fine aerosols while the dynamic approach is used for coarse aerosols.
\end{itemize}

\subsection{The aerosol modules in MICS models}

The main differences of the different aerosol modules used in the MICS models can be classified as differences in the size distribution, i.e. modal versus size resolved, the number of sections or modes, and differences in physical processes, i.e. coagulation, condensation/evaporation, cloud chemistry, heterogeneous reactions. Although input data, such as boundary conditions and emissions, are common to all models in the MICS study, sea-salt emissions are not given as input data but they are parameterized in some models. 
These differences are summarized in Table~\ref{diffmodule}.

\subsection{Polair3D}

The sensitivity study is carried out with the model $M_8$, i.e. Polair3D (\citet{bs-ijep}). The aerosol module of Polair3D is fully described in \cite{debry06siream}. Aerosols are composed of black carbon, dust, five inorganics (sulfate, nitrate, ammonium, sodium, chloride) and eight organics (\citet{ed_schell_soa}). The number of sections used in the modeling is fixed by the user. The sections' diameters are log-distributed in the diameter size-range considered, $0.01\mu$m to $10\mu$m in this study. The bounds of the sections are $0.01$, $0.02$, $0.04$, $0.08$, $0.16$, $0.32$, $0.63$, $1.26$, $2.51$, $5.01$, $10.00\mu$m.
Brownian coagulation is modeled as described in \citet{bs-debry1}, and condensation/evaporation as described in 
\citet{bs-reducaer} with a moving sectional scheme. The three approaches, the full equilibrium approach, the dynamic approach or the hybrid approach, may be used. 
For cloud chemistry, the variable size resolved module (VSRM) of \citet{faheyvsrm} can be switched on or off.
Heterogeneous reactions on the surface of particles are modeled following \citet{jacob2000}
\reaction{ HO_2 -> 0.5 H_2O_2}{r1}
\reaction{ NO_2 -> 0.5 HONO + 0.5 HNO_3}{r2}
\reaction{ NO_3 -> HNO_3}{r3}
\reaction{ N_2O_5 -> 2 HNO_3}{r4}
They are considered irreversible processes with a first order reaction rate. 

The parameterizations of sea-salt emissions and vertical diffusion are preprocessed. Sea-salt emissions are computed with the parameterization of \citet{monahan} generalized to varying relative humidity following \citet{seasaltrh}. Vertical diffusion is computed with the parameterization of \citet{troen}.

\begin{table}
\begin{tabular}{ccccc}
\hline
& \bf{$M_1$} & \bf{$M_2$} & \bf{$M_3$} & \bf{$M_4$} \\
\hline
\bf{Size distribution} & 16 bins   & None & None & 8 bins  \\
                        &       [0.02; 20$\mu$m] & & & [0.5; 90$\mu$m] \\
\bf{Thermod. model} & SCAPE2 [1] & MARS [2] & None & SCAPE2 [1]   \\
\bf{Cond./evap.} & Dynamic & Full Equil. & None & Full Equil. \\
\bf{Coagulation} & Yes & No & No &  No  \\
\bf{Heter. reactions} &  [3] & No & No & No \\
\bf{Sea-salt emission} & Yes & Yes & No & Yes \\
\hline
\end{tabular}\\
\end{table}

\begin{table}
\begin{tabular}{ccccc}
\hline
& \bf{$M_5$} & \bf{$M_6$} & \bf{$M_7$} & \bf{$M_8$}  \\
\hline
\bf{Size distribution} & 4 bins  & None & 3 modes & 10 bins \\
                        & [0.1; 10$\mu$m] & & & [0.01; 10$\mu$m] \\
\bf{Thermod. model} & SCAPE2 [1]  & [6] & ISORROPIA & ISORROPIA  \\
                    &             &      & [7] & [7] \\
\bf{Cond./evap.} & Full Equil.  & Full Equil. &  Full Equil. & Full Equil.  \\
\bf{Coagulation} &  Yes & No & Yes & Yes  \\
\bf{Heter. reactions} &  [8] & No & [9]  & No  \\
\bf{Cloud chemistry} & [4] &  [10], [11]  & [4] & No  \\
\bf{Sea-salt emission} &  Yes & No & No & No \\  
\hline
\end{tabular}
\caption{Comparison of the different aerosol modules used in the different MICS models. [1]:~\citet{scape2}, [2]:~\citet{ed_saxena_hss}, [3]:~\citet{heikes-m1}, [4]:~\citet{walcek-cmaq}, [5]:~\citet{chameides-m4}, [6]:~\citet{hov-match}, [7]:~\citet{ed_isorropia}, [8]:~\citet{tang-m5}, [9]:~\citet{riemer-het-cmaq}, [10]:~\citet{berge-m6}, [11]:~\citet{mozurkewich-m6}}
\label{diffmodule}
\end{table}

\begin{table}
\begin{tabular}{cccccccc}
\hline
& \bf{$R_1$} & \bf{$R_2$} & \bf{$R_3$} & \bf{$R_4$} & \bf{$R_5$} & \bf{$R_6$} & \bf{$R_7$}  \\
\hline
\bf{Sections} & \bf{10} & \bf{3} & 10 & 10 & 10 & 10 & 10\\
\bf{Cond./evap.} & \bf{Equil.} & Equil. & \bf{Hybrid} & Equil. & Equil.  & Equil. & Equil.   \\
\bf{Coagulation} & \bf{Yes} & Yes & Yes & \bf{No} & Yes & Yes & Yes  \\
\bf{Heter. react.} & \bf{No} & No & No & No & \bf{Yes} & No & No  \\
\bf{Cloud chem.} & \bf{No} & No & No & No & No & \bf{Yes} & No  \\
\bf{Sea-salt em.} & \bf{No} & No & No & No & No & No & \bf{Yes}  \\
\hline
\end{tabular}
\caption{Summary of the different options used in the runs performed for the sensitivity study (these configurations refer to model $M_8$).}
\label{diffruns}
\end{table}

\subsection{The sensitivity study}

In this sensitivity study, only the concentrations of sulfate and nitrate are studied.
The sensitivity tests concern the size distribution, i.e. the number of sections, and physical processes, i.e. coagulation, condensation/evaporation, cloud chemistry, heterogeneous reactions and sea-salt emissions.
Sensitivity to nucleation is not tested because nucleation would involve particles of size between $0.001\mu$m to $0.01\mu$m, a size range smaller than the one considered in this study ($0.01\mu$m to $10\mu$m).
The sensitivity to whether condensation/evaporation is treated dynamically or assuming thermodynamic equilibrium is also studied. However, the sensitivity to the thermodynamic model is not tested, although it may not be negligible as shown in \citet{edgen4}. Heterogeneous reactions as described in \citet{jacob2000} are coded in Polair3D. Other models used in MICS may however take into account different heterogeneous reactions, or similar ones but with a different reaction rate.

In the default run $R_1$, ten sections are used, condensation/evaporation processes are modeled with the full equilibrium approach, coagulation is taken into account, cloud chemistry, sea-salt emissions and heterogeneous reactions are ignored. In each of the runs $R_2$ to $R_7$,  one option differs from the run $R_1$. For example, in the run $R_2$, only three sections are used. A summary of the different runs is presented in Table~\ref{diffruns}. In the run $R_3$, the hybrid approach is used with a cutoff diameter of $0.6 \mu m$: the finest 6 sections are computed with the full equilibrium approach, while the dynamic approach is used for the coarsest 4 sections.
To show that the aerosol concentrations are not only influenced by the parameterizations in the aerosol module, a run is done with a different parametrization of the vertical diffusion. The sensitivity study of \citet{mallet05uncertainty} shows a great sensitivity to the parametrization of the vertical distribution for ozone. Therefore, the parameterization of Louis is used instead of the parametrization of Troen-Mahrt in the run $R_8$. 

The sensitivity of sulfate and nitrate concentrations to the different options is now assessed by comparing monthly averaged concentrations.

\section{Monthly averaged concentrations}

\begin{figure}
  \begin{center}
     \includegraphics[angle=0,width=0.65\textwidth]{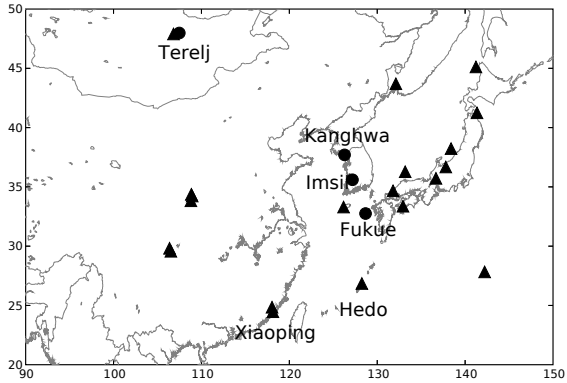}
  \end{center}
\caption{Location of the stations used in the sensitivity study}
\label{fig-station}
\end{figure}

Monthly averaged concentrations of sulfate and nitrate were measured at some of the EANET (Acid Deposition Monitoring Network in East Asia) stations \citep{mics-aer}. 
The sensitivity to the aerosol module is assessed by comparing the different runs at 23 EANET stations and at a remote site between Japan and China, Fukue, a station operated by CRIEPI \citep{mics-aer}. The location of the stations is given in Figure~\ref{fig-station}. Measurements for aerosols, that are in the computational domain of the eight models, are available at only three of the EANET stations (Terelj~--Mongolia, Kanghwa and Imsil~--Republic of Korea) and at Fukue.

\subsection{Sensitivity to the aerosol module}

To quantify the sensitivity of the different aerosol options, the normalized mean absolute error (NMAE) between $R_1$ and $R_i$ ($i=2,..., 8$) and the normalized mean bias (NMB) are computed for sulfate and nitrate, as follows (\cite{yu-metric})
\begin{eqnarray}
NMAE & = & \frac{\sum_{k=1}^N \left| C_{i,k} - C_{1,k} \right| }{\sum_{k=1}^N C_{1,k}} \; . \; 100 \% \nonumber \\
NMB & = & \frac{\sum_{k=1}^N \left( C_{i,k} - C_{1,k} \right) }{\sum_{k=1}^N C_{1,k}} \; . \; 100 \%
\end{eqnarray}
where $N$ is the number of stations (23 EANET stations + Fukue), $C_{i,k}$ ($i = 1, \; ..., 8$, $k = 1, \; ..., N$) represents the concentration of sulfate or nitrate at the station $k$ for the run $R_i$. For each run $R_i$, the higher the NMAE and NMB are, the more sensitive the results are to the option $i$.
A summary can be found in Table~\ref{stateanet}. 

\begin{table}
\begin{tabular}{ccccc}
\hline
& \multicolumn{2}{c} {\textbf{Sulfate}} & \multicolumn{2}{c} {\textbf{Nitrate}}   \\
\hline
& \bf{NMAE} & \bf{NMB} & \bf{NMAE} & \bf{NMB} \\
\hline
$R_2$ & 6\%  & 5\%   &  15\% & 15\%  \\
$R_3$ & 2\%  & 2\%   & 13\% & -12\%  \\
$R_4$ & 11\% & -10\%  & 15\% & -15\%  \\
$R_5$ & 15\% & -14\% & 99\% & 99\%  \\
$R_6$ & 17\% & 17\%  & 2\%  & -1\%  \\
$R_7$ & 3\%  & 3\%   & 10\% & 10\%  \\
$R_8$ & 34\% & 34\%  & 71\% & 71\% \\
\hline
\end{tabular}
\caption{NMAE and NMB between $R_1$ and $R_i$ ($i=2,..., 8$), for monthly averaged concentrations at 23 EANET stations and at Fukue.}
\label{stateanet}
\end{table}

As shown in Table~\ref{stateanet}, sulfate concentrations are little sensitive to options $3$ and $7$, ie. to the hybrid option and to sea-salt emissions, with a NMAE under $2\%$.
Sulfate is little sensitive to sea-salt emissions, because sea-salt sulfate is not considered in the sea-salt emissions here. Sea-salt emissions are assumed to be made exclusively of chloride and sodium.

\begin{figure}
  \begin{center}
     \includegraphics[angle=-90,width=0.55\textwidth]{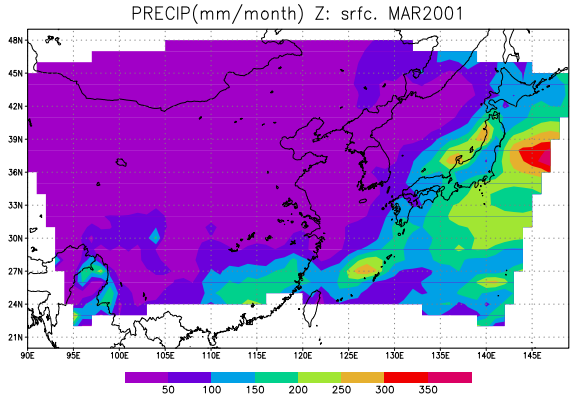}
  \end{center}
\caption{Precipitations in mm/month for March 2001.}
\label{rain}
\end{figure}

Higher sensitivity of sulfate is observed for options $2$ and $4$, i.e. for the size distribution and for coagulation with a NMAE of $6\%$ and $11\%$ respectively.
Sulfate concentrations are equally sensitive to options $5$ and $6$, i.e. to heterogeneous reactions and cloud chemistry, with a NMAE around $15\%$-$17\%$. 
Although the formation of sulfate is often thought to be dominated by aqueous production (e.g. \cite{edseinfeld}), the sensitivity of sulfate to cloud chemistry is only $17\%$ here. However, this sensitivity varies strongly with the location of stations. Figure~\ref{rain} shows the total amount of precipitations for March 2001. Precipitations are low over Mongolia and central China, but high over south China and Japan. Accordingly, the sensitivity of sulfate to cloud chemistry is only $1\%$ at Terelj in Mongolia (see Figure~\ref{fig-station}), whereas it is as high as $71\%$ and $62\%$ in Xiaoping and Hedo. Note that higher sensitivity to cloud chemistry would have been observed in July, when precipations are higher than in March.

NMAE tends to be higher for nitrate than for sulfate concentrations, except for the option 6, i.e. for cloud chemistry. Nitrate is more sensitive to the options of the aerosol module than sulfate, explaining the larger discrepancies observed between the models $M_1$, ...,$M_8$ for nitrate than for sulfate. The effects of  the size distribution, the hybrid scheme and sea-salt emissions are stronger on nitrate than sulfate because they alter thermodynamic equilibrium. Sulfate is less affected because of its low volatility.

For nitrate, the highest sensitivity is observed for option $5$, i.e. heterogeneous reactions, with a NMAE as high as $99\%$. Only cloud chemistry (option $6$) does not significantly modify nitrate concentrations. 
The effects of the size distribution, the hybrid scheme, coagulation and sea-salt emissions (options $2$, $3$, $4$ and $7$) are of similar strength in terms of NMAE with a NMAE around $10\%$-$15\%$. 

The positive NMB of the run $R_6$ for sulfate illustrates that the concentration of sulfate increases due to cloud chemistry as expected by the oxidation of dissolved $SO_2$ into sulfate.

Nitrate concentrations increase by heterogeneous reactions as shown by the high positive NMB of $99\%$ for the run $R_5$, whereas sulfate concentrations decrease although the NMB $-14\%$ is not as high in absolute value as for nitrate.
As deduced from the set of reactions \ref{reac:r1}, \ref{reac:r2}, \ref{reac:r3} and \ref{reac:r4}, heterogeneous reactions lead to higher $HNO_3$ and $H_2O_2$ concentrations and to lower $HO_2$ concentrations.  
Higher $H_2O_2$ concentrations in $R_5$ would increase sulfate concentrations in the real atmosphere. However, this does not occur in the model runs because $R_5$ excludes aqueous chemistry.
A more detailed comparison of the runs $R_1$ and $R_5$ shows that monthly averaged $OH$ and $O_3$ concentrations are also lower when heterogeneous reactions are taken into account. Because of lower $OH$ concentrations, oxidation of $SO_2$ by $OH$ is lower resulting in lower sulfate concentrations.
Lower sulfate may in turn lead to higher nitrate, which is then required to neutralize ammonia. Higher nitrate concentrations may also be a consequence of the higher $HNO_3$ concentrations that condense on particles to form ammonium-nitrate.

\begin{figure}
  \begin{center}
     \includegraphics[angle=0,width=0.49\textwidth]{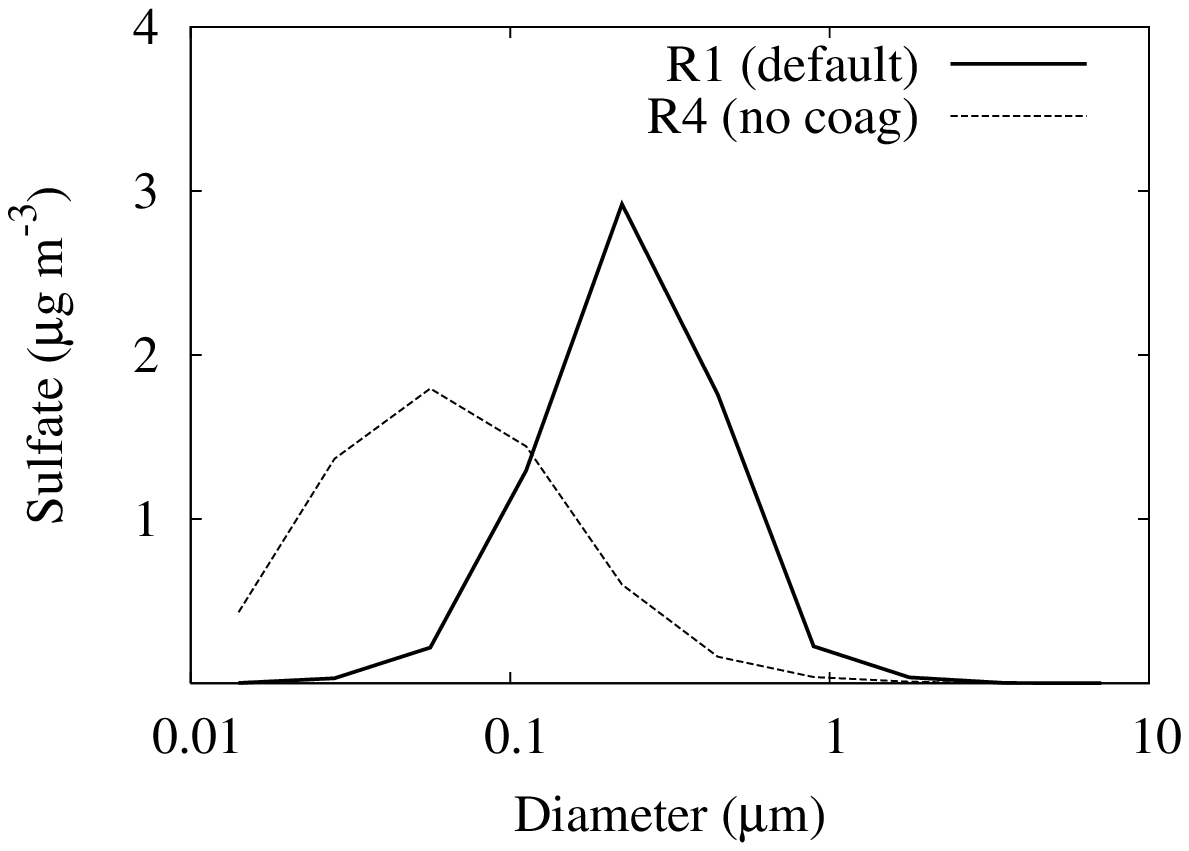}
     \includegraphics[angle=0,width=0.49\textwidth]{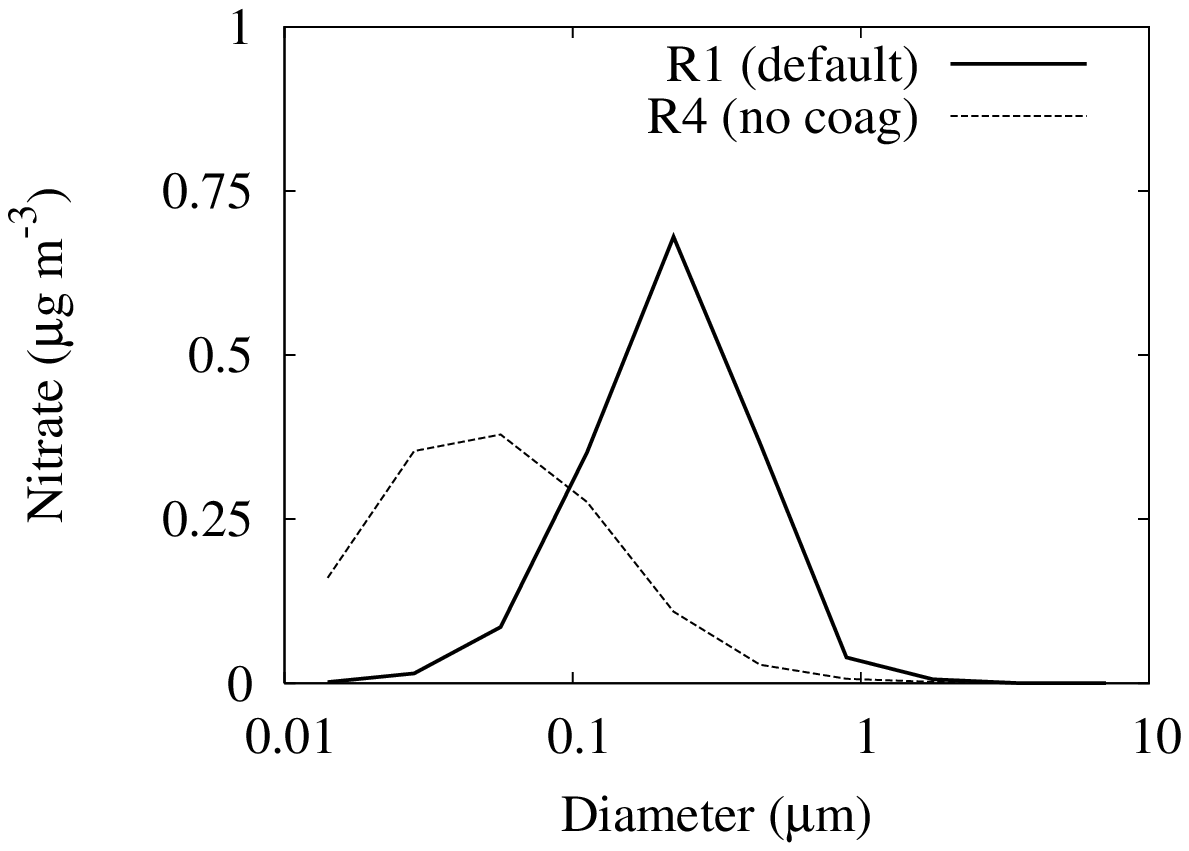}
  \end{center}
\caption{Size distribution of sulfate and nitrate monthly averaged concentrations averaged over all stations.}
\label{fig1:sizedistr}
\end{figure}

Coagulation is more efficient for small particles, limiting their concentrations. Figure~\ref{fig1:sizedistr} shows the size distribution of sulfate and nitrate monthly averaged concentrations, averaged over all stations. When coagulation is not taken into account, the distribution is centered at smaller diameters than when coagulation is taken into account. 

For both sulfate and nitrate, a high sensitivity to the parameterization of the vertical diffusion (run $R_8$) is observed. For sulfate, this sensitivity is much higher than the sensitivity to any of the options tested in the aerosol module. For nitrate, this sensitivity is very high as well, although not as high as the sensitivity to heterogeneous reactions.

 \subsection{Comparison to the sensitivity to the chemistry transport model}

To compare the sensitivity to the aerosol module to the sensitivity to the chemistry transport model, the NMAE is computed for stations where measurements are available (Terelj, Kanghwa, Imsil, Fukue) as follows
\begin{eqnarray}
NMAE & = & \frac{\sum_{k=1}^M \left| C_{i,k} - O_{k} \right| }{\sum_{k=1}^M O_{k}} \; . \; 100 \% 
\end{eqnarray}
where $M = 4$ is the number of stations where measurements are available, $O_k$ ($k = 1, \; ..., N$) is the concentration of sulfate or nitrate at station $k$. Table~\ref{mngeeanet} shows the mean NMAE for runs $R_1$ to $R_7$ as well as the smallest (Min) and largest (Max) value of NMAE obtained for these runs. These values Min, Max and mean of NMAE are computed not only for runs $R_1$ to $R_7$ but also for models $M_1$ to $M_8$, in order to compare the sensitivity of the aerosol module (runs $R_1$ to $R_7$) to the sensitivity of the chemistry transport model (models $M_1$ to $M_8$).
For sulfate, all eight models are included when computing the mean NMAE. For nitrate, the model $M_3$ is not included because it did not provide results. Furthermore, because the NMAE observed with the model $M_4$ is one order of magnitude higher than other models for nitrate, results with and without taking $M_4$ into account are presented.

\begin{table}
\begin{tabular}{ccccccc}
\hline
& \multicolumn{3}{c} {\textbf{Sulfate}} & \multicolumn{3}{c} {\textbf{Nitrate}}   \\
\hline
& \bf{Min} & \bf{Mean} & \bf{Max} & \bf{Min} & \bf{Mean} & \bf{Max} \\
$R_1-R_7$ & 12\% & 13\% & 14\% & 20\% & 34\% & 85\% \\
$M_1-M_8$ & 10\% & 14\% & 18\% & 13\% &  57\% & 188\%  \\
$M_1-M_8$ (no $M_4$) & -- & -- & --- & 13\% &  35\% & 69\%  \\
\hline
\end{tabular}
\caption{Min, mean and Max of NMAE associated to runs $R_1$ to $R_7$ and to models $M_1$ to $M_8$ for monthly averaged concentrations.}
\label{mngeeanet}
\end{table}

Values of NMAE are higher for nitrate than sulfate, confirming again the higher sensitivity of nitrate.

For sulfate, variability is larger for models $M_1$ to $M_8$ than for runs $R_1$ to $R_7$. The NMAE varies by only $2\%$ for runs $R_1$ to $R_7$ whereas it varies by as much as $8\%$ for models $M_1$ to $M_8$. For nitrate, the NMAE varies by $65\%$ for runs $R_1$ to $R_7$ whereas it varies by as much as $175\%$ for models $M_1$ to $M_8$. However, without taking into account the model $M_4$ for nitrate, the NMAE varies by $56\%$ for models $M_1$ to $M_8$, that is the variations are of the same order of magnitude as those of runs $R_1$ to $R_7$. In other words, for nitrate, the variability within the aerosol module ($R_1$ to $R_7$) is as large as the variability between models ($M_1$ to $M_8$).

Apart from the uncertainty in aerosol parameterizations, the variability among models $M_1$ to $M_8$ is linked to numerical schemes and parameterizations used for transport, diffusion, deposition, scavenging, chemical mechanism, choice of computational domain and input data. Although input data are the same for all models for emission and boundary conditions, meteorological data that are different from the standard MICS meteorological file have sometimes been used. For example, a difference in the relative humidity field used by models as high as $20\%$ has been observed \citep{hozumi}. Such variability in relative humidity contributes to variability in aerosol concentrations. 

Variability linked to numerical schemes and parameterizations outside the aerosol module is important especially for sulfate for which the variability to parameterizations in the aerosol module is not so high. For example, the sensitivity to the parameterization of the vertical diffusion has shown to be higher than the sensitivity to any options of the aerosol module for sulfate.

Because only uncertainties in aerosol parameterizations are considered for the runs $R_1$ to $R_7$, uncertainties observed for models $M_1$ to $M_8$ should be larger. This is true for sulfate. However, for nitrate, if the results of the model $M_4$ are not considered, the NMAE varies as much for models $M_1$ to $M_8$ as for runs $R_1$ to $R_7$. This strong variability of nitrate to the aerosol module stresses the difficulties in modeling nitrate concentrations accurately.

\section{Conclusion}

For March 2001, the sensitivity to the aerosol module of the model $M_8$, Polair3D, is assessed and compared to the sensitivity to the chemistry transport model.
Because in the MICS comparison, all models are assumed to use the same input data, the variability between the different models is due to differences in physical parameterizations, differences in numerical schemes and differences in the chemical mechanism. 
Concerning the sensitivity to the aerosol module, only the sensitivity to the size distribution (number of sections) and the sensitivity to physical parameterizations (coagulation, whether condensation is solved assuming full equilibrium or using an hybrid scheme, cloud chemistry, heterogeneous reactions, and sea-salt emissions) are considered. However, there is also a sensitivity to numerical algorithms for simulation of condensation/evaporation for example as shown by \citet{zhang-madrid}.

To assess the sensitivity to the aerosol module and to the chemistry transport model, monthly averaged concentrations over 24 stations are computed. Sulfate concentrations are little sensitive to the hybrid scheme and to sea-salt emissions, with a normalized mean absolute error ($NMAE$) and a normalized mean bias under $2\%$. Nitrate concentrations are little sensitive to cloud chemistry.
For sulfate, the sensitivity to cloud chemistry is important although not dominant with a NMAE of about $15\%$.
For sulfate and especially for nitrate, high sensitivity to heterogeneous reactions is observed. 
For nitrate the sensitivity to heterogeneous reactions is particularly high, with a NMAE as high as $99\%$ for monthly averaged concentrations. 

To compare the sensitivity to the chemistry transport model to the sensitivity to the aerosol module, the minimum, the mean and the maximum of the NMAE are computed for $R_i$, i.e. for runs associated to sensitivity tests on the aerosol module, and for $M_i$, i.e. for models participating in the MICS comparison. The values of NMAE are higher for nitrate than sulfate, confirming the higher sensitivity of nitrate.
For monthly averaged concentrations at all EANET stations for nitrate, the variations of the NMAE for $R_i$ are of the same order as the variations for $M_i$, suggesting a very high sensitivity to the aerosol module. For sulfate, the variations of the NMAE for $R_i$ are much smaller than the variations of the NMAE for $M_i$ ($2\%$ against $8\%$), suggesting that the sensitivity to the aerosol module is less strong than the sensitivity to the chemistry transport model. However, if $R_8$, i.e. the sensitivity to the parameterization of the vertical diffusion, is added in the computation of the variations of the NMAE for $R_i$, then the variations of the NMAE for $R_i$ are almost of the same order as the variations for $M_i$ ($6\%$ against $8\%$). 

The high variability observed for nitrate stresses the difficulties in modeling nitrate concentrations accurately. Because nitrate concentrations are highly sensitive to heterogeneous reactions, better knowledge of reaction probabilities for example, some of which may be expressed as a function of aerosol composition, temperature and relative humidity (e.g. \citet{evans-n2o5} for \ref{reac:r4}), may help to improve the models accuracy.

\section{Acknowledgements}
The authors are very thankfull to C. Bennet, G.R. Carmichael, M. Engardt, C. Fung, Z. Han, M. Kajino, U. Park, T. Sakurai, N. Thongboonchoo, H. Ueda for providing the data for the different models, the EANET monitoring data and for very valuable discussions. We would also like to thank T. Holloway for providing boundary counditions, and A. Kannari for emissions.




\bibliographystyle{elsart-harv}
\bibliography{biblio-pole}

\end{document}